\documentclass[11pt,a4paper]{article}
\pdfoutput=1

\usepackage{jcappub}

\usepackage{graphicx}

\usepackage{float}
\usepackage{dcolumn}
\usepackage{bm}
\usepackage{enumitem}
  \usepackage{tabularx}
  \usepackage{amsmath}
    \usepackage{cases}
   \newcolumntype{C}{>{\centering\arraybackslash}X}
   \newcolumntype{L}{>{\raggedright\arraybackslash}X}
   \newcolumntype{R}{>{\raggedleft\arraybackslash}X}
\setlistdepth{10}

\DeclareMathOperator{\sinc}{sinc}

\def\bea{\begin{eqnarray}}
\def\eea{\end{eqnarray}}
\def\be{\begin{equation}}
\def\ee{\end{equation}}
\def\ba{\begin{array}}
\def\ea{\end{array}}

\def\fnl{{f_{\rm NL}}}

\def\C{{\mathcal C}}

\def\exp{\textrm{exp}}

\title{PBH in single field inflation: the effect of shape dispersion and 
non-Gaussianities.}

\author{Vicente Atal, Judith Cid, Albert Escriv\`a and Jaume Garriga}
\emailAdd{
vicente.atal@icc.ub.edu, jcidgime7@alumnes.ub.edu, albert.escriva@fqa.ub.edu, jaume.garriga@ub.edu}
\affiliation{Departament de F\'isica Qu\`antica i Astrofi\'sica, i Institut  de  Ci\`encies  del  Cosmos, Universitat de Barcelona, Mart\'i i Franqu\'es 1, 08028 Barcelona, Spain.}

\abstract{Primordial black holes (PBHs) may result from  high peaks in a random field of cosmological perturbations. In single field inflationary models, such perturbations can be seeded as the inflaton overshoots a small barrier on its way down the potential. PBHs are then produced  through two distinct mechanisms, during the radiation era. The first one is the familiar collapse of large adiabatic overdensities. The second one is the collapse induced by relic bubbles where the inflaton field is trapped in a false vacuum. The latter are due to rare backward  fluctuations of the inflaton which prevented it from overshooting the barrier in horizon sized regions. We consider (numerically and analytically) the effect of non-Gaussianities on the threshold for overdensities to collapse into a PBH. Since typical high peaks have some dispersion in their shape or profile, we also consider the effect of such dispersion on the corresponding threshold for collapse. With these results we estimate the most likely channel for PBH production as a function of the non-Gaussianity parameter $\fnl$. We also compare the threshold for collapse coming from the perturbative versus the non perturbative template for the non-Gaussianity arising in this model. We show that i) for $\fnl\gtrsim 3.5$, the population of PBH coming from false vacuum regions dominates over that which comes from the collapse of large adiabatic overdensities, 
ii) the non-perturbative template of the non-Gaussianities is important to get accurate results. iii) the effect of the dispersion is small in determining the threshold for the compaction function, although it can be appreciable in determining the threshold amplitude for the curvature perturbation at low $\fnl$. We also confirm that the volume averaged compaction function provides a very accurate universal estimator for the threshold. 
}

\begin{document}

\maketitle

\section{Introduction}

Primordial Black Holes (PBH) may have formed during the radiation dominated era due to unusually high peaks in the distribution of cosmological density perturbations~\cite{Hawking:1971ei, Carr:1974nx}. There are strong observational constraints on the abundance of PBH over a wide range of mass scales \cite{Carr}. Nonetheless, these still allow for several phenomenologically interesting possibilities. For instance, PBH of sublunar  \cite{IKMTY} or stellar mass \cite{Bird,Sasaki1} may constitute a sizable fraction of all dark matter in the universe\footnote{In the case of stellar masses, the strongest constraint on the fraction $f$ of dark matter in the form of 
PBHs may come from the observed rate of merger events by the LIGO/Virgo collaboration. Nonetheless, such constraint can be substantially relaxed, or even voided, due to various environmental effects which may contribute to the eccentricity of PBH binaries at the time of their formation, or shortly after. Such effects may include 
the infall of PBH onto binaries and the collision of binaries with compact N body systems \cite{RVV,Raidal:2018bbj,Vaskonen:2019jpv}, as well as the torque exerted by an enhanced power spectrum of cosmological perturbations at small scales \cite{GT}.}. Also, the origin of supermassive black holes at the center of galaxies 
is not very well understood at present, and one possibility is that they may have formed by accretion from a smaller intermediate mass PBH seed (for a recent review, see ~\cite{Sasaki:2018dmp}).

In order to make accurate predictions on the statistical properties of PBHs, it is necessary to be specific about their formation process. One of the simplest mechanisms is the collapse of large adiabatic perturbations seeded during a period of single-field inflation \cite{Yokoyama:1998pt,GB,Kannike:2017bxn,Germani:2017bcs,Motohashi:2017kbs,Ballesteros:2017fsr,Ozsoy:2018flq,Cicoli:2018asa,Dalianis:2018frf,Bhaumik:2019tvl}. While fluctuations must be predominantly Gaussian at the cosmic microwave background scales, those leading to PBH formation at smaller scales are typically non-Gaussian \cite{Cai:2017bxr,Atal:2018neu,Passaglia:2018ixg,Atal:2019cdz}\footnote{We expect this to be the case also in other scenarios leading to PBH formation, as variants of multifield inflation \cite{Kawasaki:2012wr,Clesse:2015wea,Belotsky:2018wph} and non-canonical inflation \cite{Cai:2018tuh,Kamenshchik:2018sig,Chen:2019zza}.}. 
Sufficiently large amplification of the perturbations are induced while the inflation 
overshoots a small barrier on its way down the potential, undergoing a periond of ``constant roll" (see Fig. \ref{potential}). In this context, PBH can be formed not only from the collapse of a large adiabatic overdensity, but also from false vacuum bubbles which continue inflating in the ambient radiation dominated universe, and eventually pinch off from it. This results in a black hole which separates the ambient universe from an inflating baby universe \cite{Garriga:2015fdk,Deng:2016vzb,Atal:2019cdz}.\footnote{These are sometimes refered to as black holes with a baby universe inside. Note, however, that the baby universe is not in the trapped region, or ``interior" of the black hole. Rather, the trapped region separates two normal regions, one in the parent ambient universe and the other in the baby universe, which were once causally connected but are not anymore, after the trapped region forms.}

A question of practical interest is to determine the abundance of PBHs. Several works have already treated the influence of non-Gaussianities in the abundance of PBHs \cite{Bullock:1996at,PinaAvelino:2005rm,Hidalgo:2007vk,Young:2013oia,Young:2014ana,Young:2014oea,Young:2015cyn,Pattison:2017mbe,Atal:2019cdz,Yoo:2019pma,Kehagias:2019eil}. Since this turns to be large, it is important to i) predict the amplitude and shape of the non-Gaussianities for a given model of PBH formation, and ii) consider their influence beyond perturbation theory.

When PBHs are formed from rare overdensities, their abundance will depend on the threshold for the amplitude of the overdensity to collapse once it reenters the horizon. This threshold notoriously depends on the shape (or profile) for the overdensity \cite{Shibata:1999zs,Harada:2015yda,Germani:2018jgr,
Musco:2018rwt,Yoo:2018esr}. For a Gaussian random field, the typical shape of high peaks is determined from the power spectrum, but if the distribution is non-Gaussian, the shape will also depend on the nature of the non-Gaussianity \cite{Atal:2019cdz,Kehagias:2019eil,Yoo:2019pma}. Furthermore, since fluctuations are drawn from a statistical distribution, the shapes of perturbations susceptible of collapsing will inherit a dispersion. While the mean profile is usually taken to be representative of the typical shape, it seems important to consider how the threshold may vary due to the dispersion of shapes. This point is particularly relevant when a mean profile for the perturbations cannot be defined, as it is the case for large overdensities coming from the model of single-field inflation with a barrier\footnote{In a nutshell, the problems is that $\zeta$ diverges when the amplitude of $\zeta_g$ reaches a critical value $\mu_*$, and it is not even defined for larger amplitude of $\zeta_g$, for which there is a finite probability.}  \cite{Atal:2019cdz}.

In this work we study the dependence of the threshold on the dispersion of the profiles, including the non-Gaussianity resulting from the physics of single-field inflation. The non-Gaussianity is entirely due to the non-linear relation between the Gaussian variable 
\begin{equation}
\zeta_g \equiv -\left.H{\delta\phi\over \dot\phi}\right|_{\rm sr}, \label{Gaussian}
\end{equation}
and the non-Gaussian gauge-invariant curvature perturbation $\zeta$. Here $\delta\phi$ is the inflaton field perturbation in the flat slicing, evaluated at the onset of the slow roll attractor behaviour past the top of the barrier, and $H$ is the expansion rate during inflation\footnote{Refs. \cite{Kawasaki:2019mbl,Young:2019yug,DeLuca:2019qsy,Kehagias:2019eil} consider the non-Gaussianity in the density perturbation $\delta$ due to the non-linear relation between $\delta$ and $\zeta$. Note that such discussion would be redundant in our approach, where the initial conditions for numerical evolution, as well as the threshold estimators for gravitational collapse, are expressed directly in terms of $\zeta$.}. 
For the non-linear relation between $\zeta$ and $\zeta_g$, we will compare the non-perturbative expression which follows from the single field model where the inflaton overshoots a small barrier \cite{Atal:2019cdz}, with the more widely used perturbative Taylor expansion of $\zeta$ to second order in $\zeta_g$ (parametrized by the standard coefficient $\fnl$). These non-perturbative and perturbative versions of local non-Gaussianity are given, respectively, in Eqs. (\ref{eq:zetanptransf}) and (\ref{eq:zeta_local_trans}) below. 

We will find the thresholds for collapse into a PBH under the assumption of spherical symmetry, by using a recently developed numerical code~\cite{albert}. This solves the Misner-Sharp (MS) partial differential equations by using spectral methods. We will also compare the results obtained by numerical evolution with the results which can be obtained from a recently proposed {\em universal} estimator for the strength of  a perturbation \cite{EGS}. This is given by a suitable spatial average $\bar {\cal C}$ of the so-called initial compaction function ${\cal C}[\zeta(r)]$ \cite{Shibata:1999zs}, out to a certain optimal radius $r_m$. The threshold value for $\bar {\cal C}$ which triggers gravitational collapse turns out to be extremely robust, in the sense that it is nearly independent of the radial profile of the perturbation $\zeta(r)$. 

The plan of the paper is the following: In section \ref{sec:peaks}, we consider the typical shapes of a high peak in the curvature perturbation profile, within one standard deviation of the median profile, and we introduce the non-perturbative relation between the curvature perturbation $\zeta$ and the Gaussian variable $\zeta_g$.  In Section \ref{sec:num}, we present the Misner-Sharp equations and we review the criteria for the formation of PBHs. The results of the numerical simulation and the analytical estimates, together with their interpretation are presented in section \ref{sec:results}.

\section{Large and rare peaks from single field inflation}\label{sec:peaks}

At cosmological scales, the power spectrum of primordial perturbations must be of the order of $10^{-9}$, in accordance with observations of the cosmic microwave background. However, in order for PBH formation to be significant, the power must be of the order of $10^{-3}-10^{-2}$ at the PBH scale. This jump in the amplitude can be achieved if the inflaton field passes trough a transient period with $\ddot \phi/H\dot\phi \approx const. < -3$. Throughout this paper, we shall refer to this as ``constant-roll" (CR).\footnote{In its original definition \cite{Alexei}, constant-roll refers to any period where $\ddot \phi = -(3+\alpha) H\dot\phi$, with any constant value of $\alpha$.  Ultra slow-roll (USR) corresponds to $\alpha=0$, and can also enhance the amplitude of the power spectrum. However, to our knowledge, there is no concrete model of transient  USR where the amplification is sufficient to provide a significant abundance of PBH \cite{Atal:2018neu}. Hence, here we consider a transient period with $\alpha>0$. This corresponds to the presence of a small barrier in the potential which slows down the motion of the inflaton for a short period of time (see Fig. \ref{potential}).}


Parametrically, the fraction of dark matter in PBH is $\Omega_{PBH} \sim 10^{9} (M_\odot/M_{PBH})^{1/2}\beta_0 $, where the probability of PBH formation at the time when a large perturbation crosses the horizon can be roughly estimated as $\beta_0 \sim \exp[-\zeta_{th}^2/ 2\sigma_0^2]$, for some threshold value $\zeta_{th} \sim 1$. The remaining factors in the estimate of $\Omega_{PBH}$ account for the dilution of radiation relative to PBH, from the time of their formation until the time $t_{eq}$. For $M_{PBH}$ in the broad range $10^{-13}-10^{2} M_\odot$, the threshold for the perturbations to undergo gravitational collapse must be in the range $\zeta_{th} \sim (6 - 8) \sigma_0$, sizably larger than the standard deviation, in order to obtain a significant $\Omega_{PBH}\sim 1$. Because these perturbations are very rare, we can use the theory of high peaks to describe them.

\subsection{The typical high peak profiles}\label{typicalprofiles}

Since the non-Gaussian curvature perturbation $\zeta$ is a local function of the Gaussian field $\zeta_g$, let us start by reviewing the latter~\cite{Bardeen:1985tr}.
This will be the basis to describe the non-Gaussian realisations. Fluctuations of $\zeta_g$ are characterized by the power spectrum $P_\zeta (k)$,
representing the variance of the random field per logarithmic interval in $k$,
\begin{equation}
\langle \zeta_g^2 \rangle \equiv \sigma_{0}^{2}= \int \frac{dk}{k} P_\zeta(k).
\end{equation} 
Introducing the normalized two point correlation function of $\zeta_g(\vec x)$ as
\begin{equation}
    \psi(r)\equiv \dfrac{1}{\sigma_{0}^{2}}\langle \zeta_g(r)\zeta_g(0)\rangle=\dfrac{1}{\sigma_{0}^2}\int P_\zeta(k) \sinc{kr} \,\dfrac{dk}{k},
\end{equation}
peaks of the Gaussian random field of given amplitude $\mu= \nu \sigma_0$ at the origin, have a mean profile given by
\begin{equation}
\langle \zeta_g(r)|\nu,peak \rangle = \sigma_0 [\nu\psi(r) + O(\nu^{-1})], \label{mean}
\end{equation}
where the last term can be neglected in the limit of high peaks $\nu\gg 1$. Note that $\psi(0)=1$. The above expectation is calculated by using the number density distribution of peaks. This distribution is almost Gaussian, except for a Jacobian prefactor which relates the condition of being an extremum with the condition for the peak to be at a certain location. 
If we simply condition the field value to be at a certain height, the distribution is Gaussian, an leads to the simpler expression ~\cite{Bardeen:1985tr}
\begin{equation}
\langle \zeta_g(r)|\nu \rangle = \sigma_0 \nu\psi(r), \label{median}
\end{equation}
which coincides with the large $\nu$ limit of (\ref{mean}). For a Gaussian distribution, the mean and the median coincide, and therefore for the rest of this paper 
we shall refer to (\ref{median}) as the {\em median} Gaussian profile.

Still, there will be some deviations around the median,
so that the typical profile will be of the form
\begin{equation}
    \zeta_g(r)=\mu \psi(r) \pm \Delta \zeta, \label{zetarange}
\end{equation}
where the variance of the  shape is given by \cite{Bardeen:1985tr}
\be\label{eq:deltagaus}
\frac{\left(\Delta \zeta (r)\right)^2}{\sigma_0^2}=1-\frac{\psi^2}{1-\gamma^2}-\frac{1}{\gamma^2\left(1-\gamma^2\right)}\left(2\gamma^2\psi+\frac{R_s^2\nabla^2\psi}{3}\right)\frac{R_s^2\nabla^2\psi}{3}-\frac{5R_s^4}{\gamma^2}\left(\frac{\psi'}{r}-\frac{\nabla^2\psi}{3} \right)^2-R_s^2\frac{\psi'^2}{\gamma^2}  \ .
\ee
Here $\gamma\equiv\sigma_1^2/(\sigma_2\sigma_0)$, and $R_s\equiv \sqrt{3}\sigma_1/\sigma_2$, where the gradient moments of the power spectrum are given by 
\begin{equation}
\sigma_n^2 = \int k^{2n} P_\zeta(k) d\ln k. \label{sigman}
\end{equation}
In what follows, we  are going to consider two different forms for the enhancement of the power spectrum at the PBH scale.

{\em Monochromatic power spectrum:} This is simply an idealized a delta function enhancement, such that the power spectrum is given by
 \begin{equation}
     P^{\delta}_{\zeta}(k)= \sigma_0^2 k_{0} \delta(k-k_{0}). \label{deltaspectrum}
 \end{equation}
In this case the median shape in (\ref{zetarange}) is given by
 \begin{equation}
     \psi(r)=\sinc({k_{0}r}),
     \label{zeta}
 \end{equation}
while the dispersion takes the following form
\be \label{deltadelta}
\frac{\left(\Delta \zeta (r)\right)^2}{\sigma_0^2} = 1 - \psi^2 - 5\left[ R_s^2 {\psi' \over r} +\psi\right]^2 -R_s^2 (\psi')^2 \ .
\ee
Note that in this case, we have $\gamma=1$, and the general expression (\ref{eq:deltagaus}) contains indeterminate ratios. In order to obtain (\ref{deltadelta}) we have regularized
the delta function by using a normalized distribution which is constant in an interval of radius ${\varepsilon}$ around $k=k_0$, and vanishes outside of this interval, taking the limit 
${\varepsilon}\to 0$ at the end.

{\em Sharply peaked power spectrum:} We are also going to consider a more realistic case, in which the enhancement follows a power law growth $k^n$. Models of the type considered here tend to have $n$ in the range $3-4$  \cite{Byrnes:2018txb}, and for definiteness we shall consider $n=4$.\footnote{In the example considered in \cite{Atal:2019cdz} the value $n=4$ corresponds to large $\fnl$. It was argued in  \cite{Byrnes:2018txb} that this may be the maximum possible value in canonical single field scenarios. It has recently been shown, however, that a slightly steeper spectrum is possible in certain models \cite{Carrilho:2019oqg}.}
We also consider a rapid fall of the power spectrum after the peak. In the single field model, such fall-off behaves as $k^{-{12\over 5} \fnl}$ \cite{Atal:2019cdz}. Note that at low $\fnl<5/3$, the fall-off is not sharp enough to make $\sigma_2$ [given in (\ref{sigman})] indepent on the ultraviolet details of the spectrum. In other words, peak theory cannot be blindly used in this case to find the number density of PBH at the scale of the peak $k_p$, because the distribution of peaks is dominated by smaller scales. In what follows, we will simply introduce a sharp cut-off at the peak value $k_p$. This amounts to a top hat window function in momentum space, which filters out the smaller scales.\footnote{For recent discussions on the use of window functions in the present context, see e.g. \cite{Young:2019yug,Ando:2018qdb,Young:2019osy,Kalaja:2019uju}. Since we are mostly interested in the effect of non-Gaussianity, and to avoid unnecessary complication, we shall not dwell further on this interesting issue. Nonetheless, we emphasize that for $\fnl \gtrsim 2$ the use of a window function is not strictly necessary.} The spectrum is then given by
\begin{equation}
  P^{\rm{sf}}_{\zeta}(k) = 
   \begin{cases}
    0  & \text{for } k < k_{0}, \\
     P_0 \left(\frac{k}{k_{\rm p}}\right)^{4} \ , & \text{for } k_0 \leq k \leq k_{\rm p} \\
   0 \ , & \text{for } k > k_{\rm p}\ .
   \end{cases}\label{eq:model_pw}
\end{equation}
In this case, the correlation function determining the shape of the peak is given by
\be
\psi(r)\simeq\frac{4}{k_{\rm p}^4r^4}\left[-2+\left(2-k_{\rm p}^2r^2\right)\cos\left(k_{\rm p}r\right)+2k_{\rm p}r\sin\left(k_{\rm p}r\right) \right],
\ee
where we have further assumed that $k_0\ll k_{\rm p}$. For its dispersion we can take directly Eq. (\ref{eq:deltagaus}), since in this case $\gamma\neq 1$. We now discuss the effect of non-Gaussianities.

\subsection{Non-Gaussianity}\label{nongaussianity}

In single-field inflation, when the inflaton passes through a period of constant-roll as it overshoots a barrier, the non-Gaussian curvature perturbation $\zeta$ is related to the Gaussian field $\zeta_g$ defined in (\ref{Gaussian}) as \cite{Atal:2019cdz}
\be\label{eq:zetanptransf}
\zeta = -\mu_*\ln\left(1-\frac{\zeta_g}{\mu_*} \right).
\ee
The parameter $\mu_*$ can be written as a function of the potential as
\be
\frac{1}{\mu_*}=\frac{1}{2}\left(-3+\sqrt{9-12\eta}\right) \ ,
\ee
with $\eta\equiv V''/V$, evaluated at the local maximum of the barrier. The relation (\ref{eq:zetanptransf}) is only defined for perturbations with $\zeta_g<\mu_*$. 
Perturbations with $\zeta_g > \mu_*$ are so large that they prevent the inflaton field from overshooting the local maximum \cite{Atal:2019cdz}. The regions where the inflaton is trapped in the false vacuum are localized false vacuum bubbles which, from the point of view outside observers, end up forming a black hole, while from the point of view of internal observers they continue inflating. That is the reason why such PBH are said to carry a baby universe inside \cite{Garriga:2015fdk,Deng:2016vzb}. In this context, black holes can be formed in two different ways. If $\zeta_g$ is larger than a certain threshold $\mu_{th}$, whith $\mu_{th}<\zeta_g <\mu_*$, then standard black holes will be created by the gravitational collapse of the adiabatic overdensity. On the other hand, regions where $\zeta_g>\mu_*$, will lead to false vacuum bubbles.\footnote{Note that $\mu_{th}$ is always smaller than $\mu_*$. Since $\zeta'$ diverges as zeta approaches $\mu_*$, the compaction function will unavoidable be larger than its threshold for collapse for an amplitude $\mu_{th} < \mu_*$.}

\begin{figure}
\centering
\includegraphics[scale=0.4]{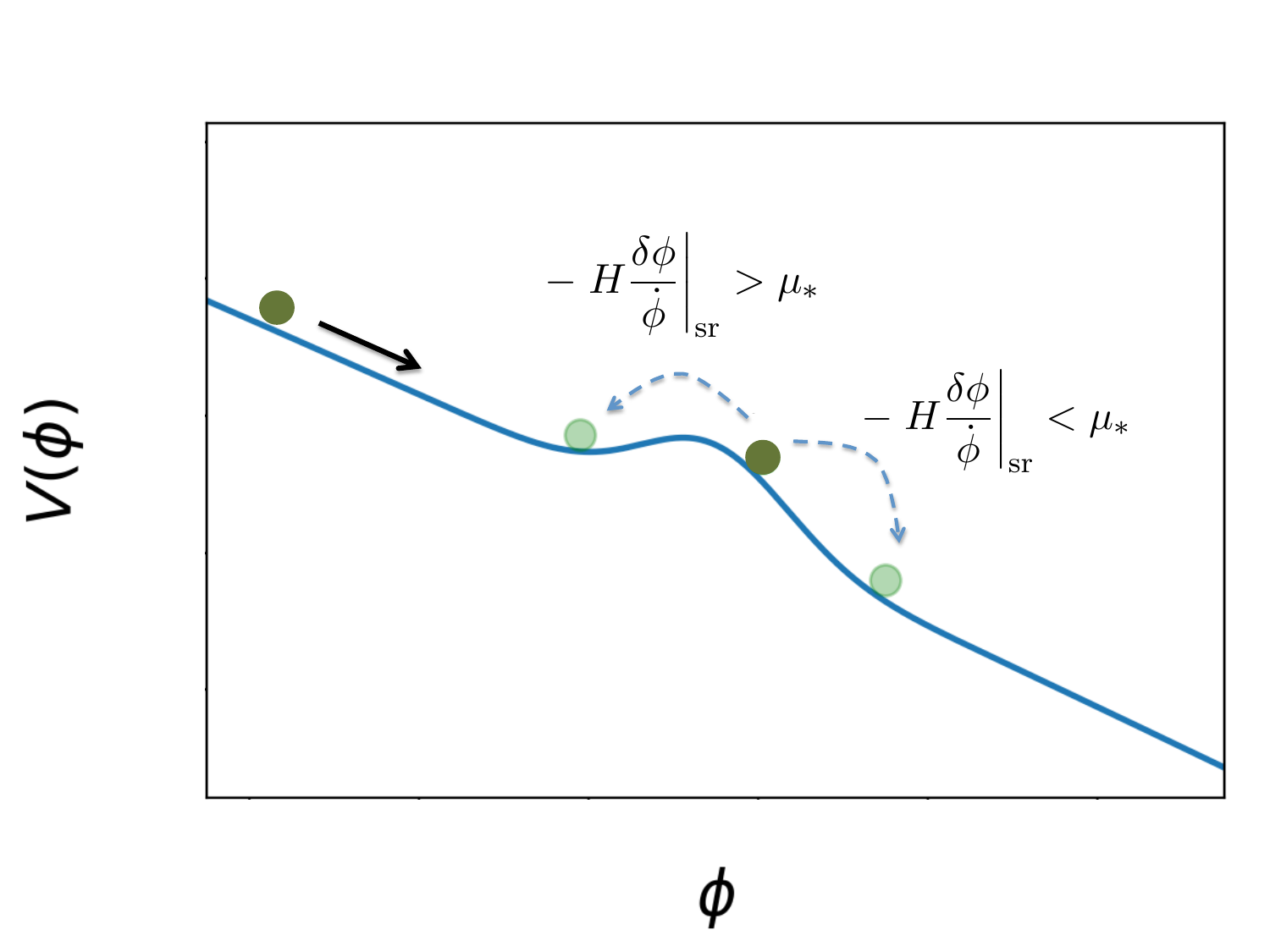}
\caption{An inflaton potential with a small barrier on its slope. As the background field goes over the barrier, it undergoes a period of constant-roll with $\ddot\phi/H\dot\phi \approx const. <-3$, which strongly amplifies the power spectrum of adiabatic perturbations to the amplitude required for significant PBH production.
Here $\delta\phi$ is the inflaton field perturbation in the flat slicing, evaluated at the onset of the slow roll attractor behaviour past the top of the barrier.
Large backward fluctuations with $-H\delta\phi/\dot\phi|_{\rm sr}>\mu_*$ may prevent some horizon sized regions from overshooting the barrier, generating false vacuum bubble relics \cite{Atal:2019cdz}. From the point of view of internal observers, these continue inflating at a high rate, while from the external point of view, these bubbles will form PBH once they enter the horizon during the radiation dominated era.\label{potential}}
\end{figure}

By Taylor expanding the non-perturbatuve relation (\ref{eq:zetanptransf}) to quadratic order in $\zeta_g$, we obtain the widely used perturbative template of local type 
non-Gaussianity,
\begin{equation}
\label{eq:zeta_local_trans}
\zeta=\zeta_g + \frac{3}{5}\fnl \zeta_{g}^2.
\end{equation}
The parameter $\mu_*$ is related to $\fnl$ through
\begin{equation}
\frac{1}{\mu_*}=\frac{6}{5}\fnl.  \label{relation}
\end{equation}
It is clear, however, that this truncated expansion is far from accurate, since PBH formation occurs in the regime where $\zeta_g$ is not small. Furthermore, the perturbative template does not capture the existence of a second channel for PBH production from regions trapped in a false vacuum, since $\zeta$ in Eq.(\ref{eq:zeta_local_trans}) is well defined for any amplitude of $\zeta_g$. Nonetheless, because of its prevalence in the literature, and in order to compare with other approaches, it seems of some interest to also consider this quadratic template.

Hence, in the following we will consider two different cases. Case A corresponds to an idealized Dirac delta function power spectrum for the Gaussian variable $\zeta_g$, as in Eq. (\ref{deltaspectrum}), where we will consider the ``vanilla" perturbative local template (\ref{eq:zeta_local_trans}) in order to obtain the non-Gaussian curvature perturbation.
Case B is a more realistic scenario based on the single field model of \cite{Atal:2019cdz}, where the logarithmic template for non-Gaussianity will be combined with the power spectrum (\ref{eq:model_pw}) in order to determine the range of typical shapes for $\zeta$.

\section{The formation of PBH}\label{sec:num}

In this section we describe the relevant equations for the evolution of spherically symmetric perturbations, which can be solved with the help of the numerical code recently developed in \cite{albert}. We also describe the criteria for creation of a BH.

\subsection{The Misner-Sharp equations}
The Misner-Sharp equations (MS) are the Einstein's equations for a spherically symmetric spacetime, in a frame comoving with a perfect fluid~\cite{MS}. In this case, the metric can be written in the general diagonal form
\begin{equation}\label{metric1}
ds^2=-A(r,t)^2 dt^2+B(r,t)^2dr^2 +R(r,t)^2d\Omega^2
\end{equation}
where $d\Omega=d\theta^2 +\sin^2{\theta}d\phi^2$ is the metric on the unit 2-sphere. The fluid is at rest relative to the radial coordinate $r$. The metric components $A(r,t)$, $B(r,t)$ and $R(r,t)$ are all positive, the latter one corresponding to the areal radius of the 2-spheres. Following~\cite{MS}, we define the partial derivatives with respect to 
proper time and proper distance as
\begin{gather}
    D_{t}\equiv \dfrac{1}{A}\frac{\partial}{\partial t} \,\,\,\,\,\,\,\,\, \text{and} \,\,\,\,\,\,\,\,\, D_{r}\equiv \dfrac{1}{B}\frac{\partial}{\partial r}.
\end{gather}
Applying the last two operators to $R$ we can also define
\begin{gather}
    U\equiv D_{t}R=\dfrac{1}{A}\frac{\partial R}{\partial t} \label{DR}, \\
    \Gamma \equiv D_{r}R\equiv \dfrac{1}{B}\frac{\partial R}{\partial r}.\label{RP}
\end{gather}
We may now introduce the Misner-Sharp mass $M(r,t)$ through the equation
\begin{equation}
    \Gamma^2-U^2=1-\dfrac{2M}{R}.
    \label{gamma2}
\end{equation}
We will also use the form of the stress energy tensor of a perfect fluid
\begin{equation}
    T^{\mu\nu}=(p+\rho)u^{\mu}u^{\nu}+pg^{\mu\nu},
    \label{emtensor}
\end{equation}
where $u^{\mu}=(A^{-1},0,0,0)$ is the four-velocity field of the fluid, $\rho$ is the energy density, and $p$ is the pressure. During the radiation era, the equation of state is $p=\frac{1}{3}\rho$. In terms of these variables, the MS equations take the form
\begin{gather}
    D_{t}U=-\dfrac{\Gamma}{\rho + p}D_{r}p-\dfrac{M}{R^2}-4\pi R p, \label{DU}\\ 
    D_{t}\rho=-\dfrac{\rho+p}{\Gamma R^2}D_{r}(R^{2}U),\label{Drho}\\
    D_{r}A=-\dfrac{A}{\rho + p}D_{r}p, \label{DA} \\
    D_{t}M = -4 \pi R^{2} U p, \label{lewis}\\
    D_{r}M=4\pi R^{2}\Gamma \rho. \label{DM}
\end{gather}
Eq. (\ref{DA}) can readily be solved to obtain 
\begin{equation}
A(r,t)=[\rho_b(t)/\rho(r,t)]^{1/4}, 
\end{equation}
where we have imposed the boundary condition that $A\to 1$ at large distance from the origin, so that $t$ becomes the proper time of a homogeneous radiation dominated FLRW universe with density $\rho_b(t)$. Then, from (\ref{DU}), (\ref{Drho}), (\ref{DM}), and
(\ref{DR}) one obtains a closed set of equations for the time evolution of the variables $(U,\rho,M,R)$, after eliminating $\Gamma$ by using (\ref{gamma2}), and $B$ by using (\ref{RP}). Eq. (\ref{lewis}) is the Hamiltonian constraint, a redundant equation which is useful in order to check the accuracy of the time evolution. 

Let us now discuss the initial conditions for evolution in terms of the random field $\zeta(r)$ of primordial curvature perturbations.

\subsection{The long wavelength approximation and initial conditions}

Initially, at early times, perturbations have a physical wavelength $L$ much larger than the Hubble radius $H^{-1}$ \cite{Shibata:1999zs}. Hence, we are going to consider the long wavelength approximation to determine the form of our initial metric and hydrodynamical variables. This is based in expanding the exact solutions in a power series of a parameter 
\begin{equation}
    \epsilon(t) \equiv \dfrac{1}{H(t)L(t)},
\end{equation}
to the lowest non-vanishing order in $\epsilon(t) \ll 1$. In the limit $\epsilon\to 0$, the metric of a perturbed FRW model can be written in the form~
\begin{equation}
    ds^2=-dt^2+a^2(t)e^{2\zeta(r)}(dr^2 + r^2 d\Omega^2).
    \label{metric2}
\end{equation}
This is in a coordinate system where the energy density of the fluid is used as a clock, so that $t=const.$ surfaces coincide with $\rho=const.$ surfaces. He have also restricted to spherical symmetry, which excludes the presence of tensor modes (gravitational waves).
In terms of $\zeta(r)$, the long wavelength solution of the MS equations reads \cite{Musco:2018rwt}
\begin{gather*}
     U=H(t)R(1+\epsilon^{2}\tilde{U}),\\
    \rho=\rho_{b}(1+\epsilon^{2}\tilde{\rho}),\\
    M=\dfrac{4\pi}{3}\rho_{b} R^{3}(1+\epsilon^{2}\tilde{M})=\dfrac{4\pi}{3}\rho_{b} R^{3}(1-4\epsilon^{2}\tilde{U}),\\
     R=a(t)e^{\zeta(r)}r(1+\epsilon^{2}\tilde{R})=a(t)e^{\zeta(r)}r\left(1-\epsilon^{2}\dfrac{\tilde{\rho}}{8}+\epsilon^{2}\dfrac{\tilde{U}}{2}\right),
\end{gather*}
where the functions $\tilde{\rho}$, $\tilde{U}$ represent the energy density and velocity perturbation, given by 
\begin{gather*}
       \tilde{U}=-\dfrac{1}{6}\dfrac{e^{2\zeta(r_{k})}}{e^{2\zeta(r)}}\zeta'(r)\left[\dfrac{2}{r}+\zeta'(r)\right]r_{k}^{2},\\
        \tilde{\rho}=-\dfrac{4}{9}\dfrac{e^{2\zeta(r_{k})}}{e^{2\zeta(r)}}r_{k}^{2}\biggl[\zeta''(r) +\zeta'(r)
    \biggl({2\over r}+ {\zeta' \over 2}\biggr)
    \biggr].
\end{gather*}
Here $r_k$ is the comoving lengthscale of the perturbation associated to the wavenumber k, i.e. $r_{k}e^{\zeta(r_{k})}=[H(t)a(t)\epsilon]^{-1}$. As mentioned below Eq. (\ref{DM}), all remaining variables can be obtained from the set $(U,\rho,M,R)$, and so it is not necessary to specify any additional initial conditions.

\subsection{The criterion for BH production}

The formation of a black hole for a given initial condition can be inferred from the behaviour of perturbations which do not dissipate after entering the horizon but continue growing until a {\em trapped surface}~\cite{penrose} is formed. This signals the onset of gravitational collapse. To identify the trapped surfaces, we consider the expansion  $\Theta^{\pm} \equiv h^{\mu\nu} \nabla_{\mu}k_{\nu}^{\pm}$ of null geodesic congruences $k^{\pm}_{\mu}$ orthogonal to a spherical surface $\Sigma$. Here $h_{\mu\nu}$ is  the metric induced on $\Sigma$. There are two such congruences, which we may call inward and outward directed, with components $k_{\mu}^{\pm} = (A,\pm B,0,0)$, such that $k^+\cdot k^- = -2$. In flat space, $\Theta^{-}<0$, while $\Theta^{+}>0$. Surfaces $\Sigma$ with this property are called ``normal". If both expansions are negative, the surface is called ``trapped", while if both are positive, the surface is ``anti-trapped". In terms of the MS variables~\cite{trappedsm}, we have
\begin{gather}
    \Theta^{\pm}=\dfrac{2}{R}(U\pm \Gamma).
\end{gather}
In a spherically symmetric spacetime, any point in the $(r,t)$ plane can be thought of as a closed surface $\Sigma$ of proper radius $R(r,t)$. We can classify such points into normal, trapped and anti-trapped.
In the transition from a normal region to a trapped region, we must go through a boundary where $\Theta^-<0$ and $\Theta^+=0$. This is a marginally trapped surface which is usually called the apparent horizon. Since $\Theta^{+}\Theta^{-}\propto U^{2}-\Gamma^{2}=0$ and using Eq.~(\ref{gamma2}), the condition for the formation of an apparent horizon is simply
\begin{equation}
    R=2M.
    \label{trapped surface condition}
\end{equation}
This could be marginally trapped, as it occurs for black holes, or marginally anti-trapped, as is the case for a cosmological horizon.
If the condition $R<2M$ is satisfied in the vicinity of the apparent horizon, this means that we have trapped surfaces, and a PBH will be formed in the subsequent evolution.

A useful estimator for the strength of a spherically symmetric perturbation is the so-called compaction function, which is the mass excess $\delta M(R) = M-M_b$ enclosed in the aereal radius $R(r,t)$ relative to the FLRW background $M_{b}$, divided by the areal radius\footnote{Here we use the definition of the compaction function given originally in \cite{Shibata:1999zs}, which has also been used in most of the subsequent literature. Note, however, that some recent papers use a convention which  differs by a factor of 2.}~\cite{Shibata:1999zs}
\begin{equation}
    \mathcal{C}(r,t)\equiv \dfrac{\delta M}{R}.
\end{equation}
From (\ref{DM}), the MS mass out to the aereal radius $R$ is given by
\begin{equation}
M(R)= \int_0^{R} \rho\, dV_{MS},
\end{equation}
and a similar expression for $M_b(R)$, where $\rho$ is replaced by the background density $\rho_b$.
Here we have introduced 
the volume element\footnote{Note that this differs from the proper volume element on $t=const.$ hypersurfaces $dV_{MS}=\Gamma dV_{proper}$.} 
\begin{equation}
dV_{MS} = 4\pi R^2 dR \approx 4 \pi a^3 (1+r\zeta') e^{3\zeta} r^2 d r,\label{MSvol}
\end{equation}
where $dR = R' dr$ is evaluated on $t=const.$ hypersurfaces. In the last step whe consider 
the long wavelength limit, which is valid at sufficiently early times, when $\zeta(r)$ is time independent and $R\approx a(t) r e^{\zeta(r)}$. It is straigthforward to check that the compaction function  can also be 
expressed as
\begin{equation}
\mathcal{C}(r,t) = {1\over 2} \bar\delta (HR)^{2},
\end{equation}
where we have introduced the ``volume" averaged density perturbation
\begin{equation}
\bar \delta = {1\over V_{MS}(R)} \int_{0}^R \delta(r,t) dV_{MS}.
\end{equation} 
Let us note that the condition $R<2M$ for the formation of a trapped surface is related the criterion $\mathcal{C}_{\rm max} \approx1/2$ suggested in \cite{albert}, where $\mathcal{C}_{\rm max}(r,t)$ is the maximum value of the compaction function at some given moment of time. Note that, indeed, for $\mathcal{C}_{\rm max} \geq1/2$ we have $2M/R = 2\mathcal{C}_{\rm max}+ 2M_{b}/R > 1$, guaranteeing that the surface is trapped. From a practical point of view, both criteria perform with similar efficiency
in the simulations we have run.

\subsection{A universal threshold for collapse}

In the long wavelength limit, the compaction function is time independent, and can be expressed in terms of the curvature perturbation as \cite{Harada:2015yda}\begin{equation}
    \mathcal{C}(r)=\dfrac{1}{3}(1-(1+r\zeta'(r))^{2}), \label{cz}
\end{equation}
where the prime denotes the partial derivative with respect to the radial coordinate. 
It has long been recognized that the initial compaction function ${\mathcal C}(r)$ is a useful tool for predicting whether a pertrubation will end up collapsing into a PBH. 
If $\mathcal{C}(r)$ has a maximum at $r=r_m$ which satisfies ${\mathcal C}(r_m) > {\mathcal C}_{th}$, then a PBH will be formed after $R(r_m,t)$ enters the horizon~\cite{Shibata:1999zs}. An interesting feature of the threshold value $C_{th}$, is that its possible range is rather limited. Indeed, 
from Eq.~(\ref{cz}) it is easy to see that ${\mathcal C}_{th}$ cannot be larger than 1/3. Also, it was argued in \cite{Harada:2015yda}
that on physical grounds ${\mathcal C}_{th}\gtrsim 0.21$. The precise value of this lower bound is not very tightly determined by the argument, but recently it has been shown by numerical studies that slightly lower values are possible~\cite{EGS}, and that the threshold lies in the range
\begin{equation}
1/5 \leq {\mathcal C}_{th}\leq 1/3, \label{limits}
\end{equation}
for a broad class of shapes. Since this window is relatively narrow, spanning less than a factor of $2$, the use of a threshold $\C_{th}$ has been popular in phenomenological studies of PBH production. Nonetheless, the precise value of $\C_{th}$ still depends on the profile of the perturbation, and so in this approach we cannot completely dispense with numerical simulations of collapse in order to obtain accurate results.


Remarkably, a universal estimator has recently been proposed, whose threshold value for PBH formation seems to be independent on the shape of the high peak overdensity
\cite{EGS}. This consists of a spatial average of the compaction function out to the optimal radius $r_m$ corresponding to the maximum of ${\mathcal C}$, where again, the MS volume element (\ref{MSvol}) is used for averaging\footnote{Note that, since $\C$ is itself a spatial average of the density perturbation, this new estimator can be thought of as a double average.} 
\be\label{eq:universal}
\bar {\mathcal C} \equiv {1\over V_{MS}(R_m)}\int_0^{R_m} \C(r) \,dV_{MS} \ ,
\ee
where $R_m=R(r_m,t)$.
The shape-independent threshold value for gravitational collapse is then given by 
\begin{equation}
\bar{\mathcal C}_{th} \approx 1/5. \label{criterion}
\end{equation}
This universal behaviour has been tested in \cite{EGS} for a very broad class of shapes. Here we shall further confirm its validity by checking that it holds to very good accuracy in the class of profiles that we will study.

\section{Results}\label{sec:results}


Here we consider the thresholds for collapse for a set of typical profiles corresponding to Case A and Case B described at the end of Subsection \ref{nongaussianity}, for values of the non-Gaussianity parameter in the range $0<\fnl<6$. Ideally, we would be interested in the set of profiles
\begin{equation}
\zeta_g =\mu \psi \pm s \Delta \zeta, 
\end{equation}
which are within $s$ standard deviations from the median profile for a given amplitude $\mu$. For $s=1$ this includes 68\% of all realizations, including generic profiles which are not spherically symmetric. Nonetheless, in the limit $\nu = \mu/\sigma_0 \gg 1$ they will be approximately spherical, with corrections of order $\nu^{-1}$ \cite{Bardeen:1985tr}. Since our numerical code assumes spherical symmetry, here we shall restrict attention to profiles with such symmetry. Aside from the median shape, $\bar\zeta(r) = \mu \psi(r)$, we shall consider the profiles 
\begin{equation}
\zeta_g^{\pm}(r) = \mu \psi(r) \pm \Delta\zeta(r), \label{deltapm}
\end{equation}
with $\Delta(r)$ given by (\ref{eq:deltagaus}). These are, roughly speaking, the envolvent of all realizations within one standard deviation from the median. Denoting by $\mu_{th}^{\pm}$ and $\C_{th}^{\pm}$ the corresponding thresholds for the amplitude and the optimized initial compaction function, the differences 
\begin{equation}
\sigma_\mu= {|\mu^+_{th}-\mu^-_{th}|\over 2},\quad \sigma_\C = {|\C^+_{th}-\C^-_{th}|\over 2}. \label{dispersions}
\end{equation}
can be taken as indicative of the dispersion in the thresholds, within one standard deviation.\footnote{Departures from spherical symmetry are expected to increase the threshold value for PBH formation \cite{Kuhnel:2016exn}. A more precise study of this effect would require the development of a numerical code which can handle deviations from spherical symmetry in the ensemble of realizations. This is beyond the scope of the present work, and is left for further research.\label{sphericalfoot} }

We have determined the thresholds by using two different methods. 
Namely, by numerical evolution with the code developed in \cite{albert}, and by using the universal criterion based on $\bar \C=1/5$. 
In Fig.~\ref{fig:thresholds1} we show the results for the thresholds $\mu_{th}$ and $C_{th}$ evaluated from these two methods, in the case where we do not include the dispersion $\Delta \zeta=0$, for the perturbative and non perturbative template. We see a good agreement between both, within a deviation of $\sim2\%$, as was reported in \cite{EGS}.  The dispersions in the thresholds given in Eq. (\ref{dispersions}) are represented in Fig.~\ref{fig:thresholds2}.
Let us now comment on the more qualitative features of the results and their physical implications.

\subsection{Case A: Perturbative template}

This case corresponds to the Dirac delta function power spectrum (\ref{deltaspectrum}), together with the perturbative local template (\ref{eq:zeta_local_trans}) for the relation between $\zeta_g$ and the curvature perturbation $\zeta$.

\begin{figure}
\centering
\includegraphics[scale=0.7]{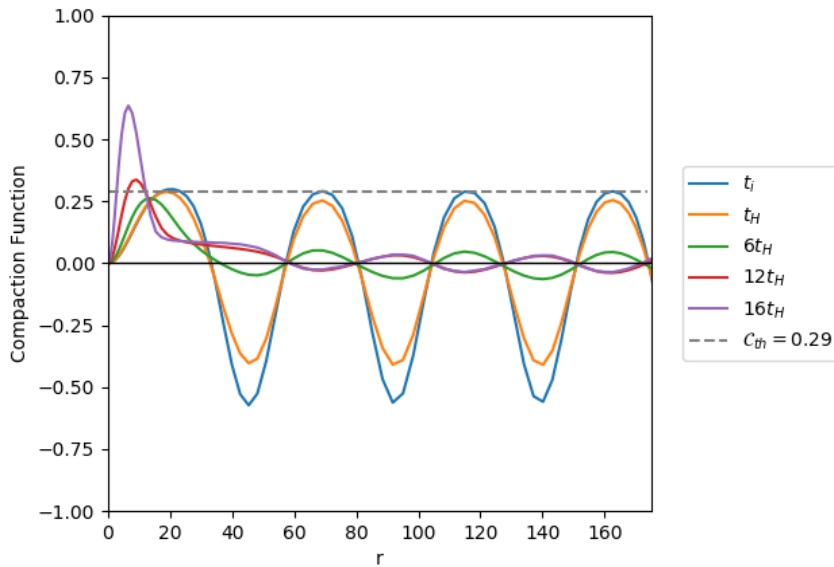}
\caption{Time evolution of the compaction function $\C (r,t)$ for the Gaussian profile (\ref{zeta}), with amplitude $\mu=0.64$, slightly larger than the threshold value $\mu_{th} \approx 0.61$. For reference, the threshold value $\C_{th}$ is indicated as a dashed line. The radial coordinate is in units of the initial time $t_i$, which we take to be much smaller than the time $t_H$ at which $r_m$ crosses the horizon, $t_H= 100\, t_i$. The size of the grid is actually somewhat larger than displayed, with $r_{max}=200\, t_i$, much larger than the initial Hubble radius $H_i^{-1}=2 t_i$. After the time $t_H$ the secondary peaks in the compaction function dissipate due to pressure gradients. The dominant peak, on the other hand, continues to grow. By the time $t=16 t_H$, the compaction function has reached values significantly larger than 1/2, indicating that a trapped region has already formed.}
\label{sincfigure}
\end{figure}

In Fig. \ref{sincfigure} we display the time evolution of the mean profile (\ref{zeta}) for the Gaussian case ($\fnl=0$). The ``sinc" profile (\ref{zeta}) is somewhat peculiar, in that the initial compaction function (represented as a blue line in the figure) has a dominant peak at $r=r_m \approx 2.7 k_0^{-1}$, and then an infinite number of nearly equally spaced secondary peaks of nearly equal height at $r\gg r_m$. The threshold for gravitational collapse of the dominant peak once it enters the horizon is determined numerically to be $\C_{th} \approx 0.29$.
This raises the somewhat naive question of what happens to the secondary peaks if the compaction function exceeds $\C_{th}$ also at the secondary peaks. Will these also trigger the gravitational collapse of bigger PBHs once they enter the horizon? It is clear from the figure that this will not be the case. As soon as the dominant peak enters the horizon, at the time $t_H$, the width of the secondary peaks will also be within the horizon, and we see that these secondary structures disipate due to pressure gradients.\footnote{The simulation is done under the assumption of spherical symmetry. However, it should be noted that for $\nu\lesssim 8$ the variance in the shapes $\zeta(r)$ at the secondary peaks is comparable to the the amplitude of $\zeta$, which means that the assumption of spherical symmetry does not really hold there. This is another reason why we do not expect these additional structures to form bigger PBHs. We thank Chulmoon Yoo for bringing this point to our attention.} By contrast, the dominant peak continues to grow and in a time-scale $t\sim 10 t_H$, it reaches $\C >1/2$, signaling the existence of a trapped region with $2M>R$.

In fact, for the profiles $\zeta_g^{\pm} $ given in (\ref{deltapm}), we find that the initial compaction function for $\zeta_g^+$ can be lower at the first peak than it is at the subsequent ``secondary" ones. Still, the first peak is the one that grows under time evolution, until a trapped surface forms, whereas the secondary ones dissipate. This is important, because it highlights the fact that the relevant optimal radius $r_m$ at which we evaluate $\C(r_m)$ in order to determine the threshold -- and which also enters the universal estimator Eq. (\ref{eq:universal}) -- is not the absolute maximum of the compaction function, but the local maximum which is closest to the origin.

\begin{figure}
\centering
\includegraphics[scale=0.31]{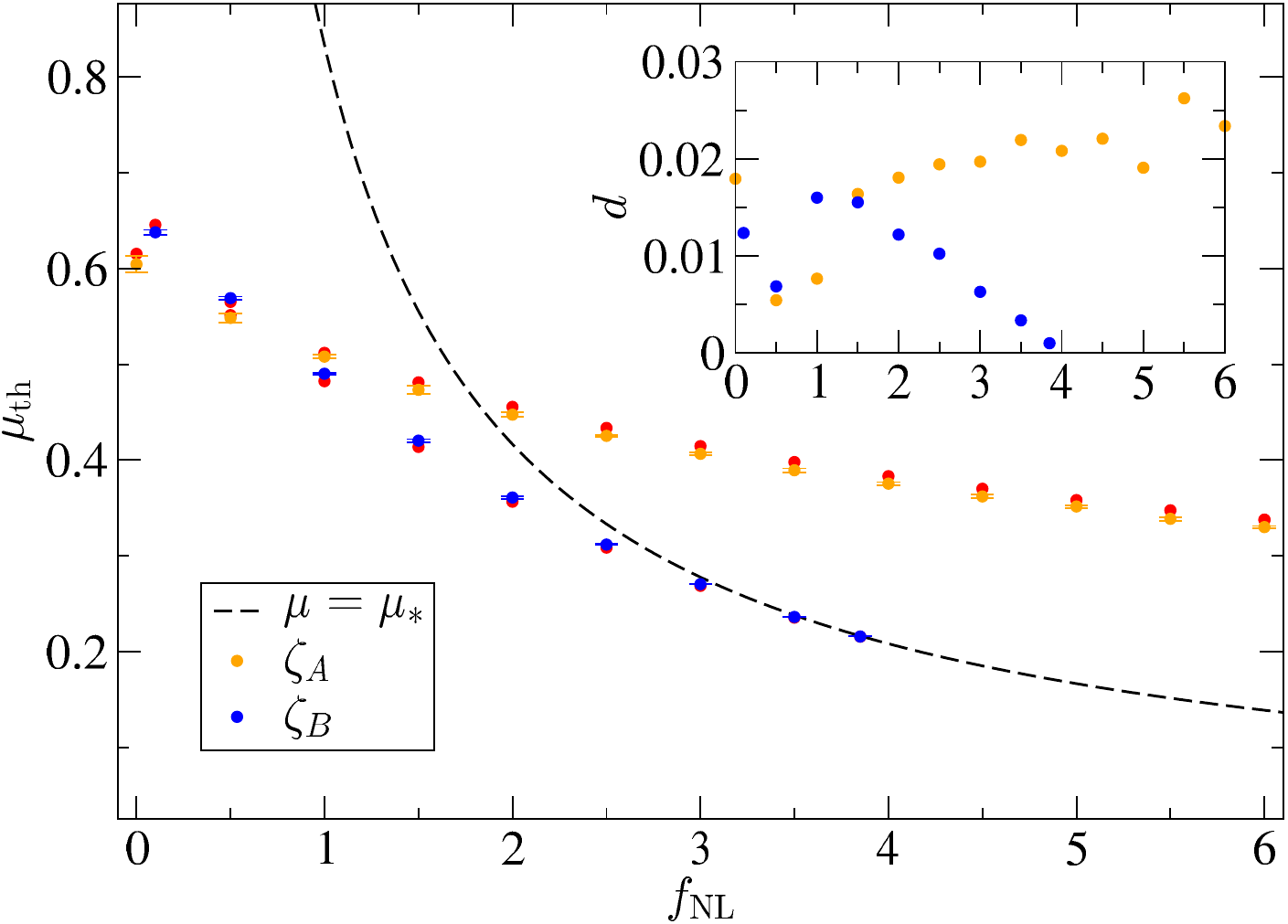}
\includegraphics[scale=0.31]{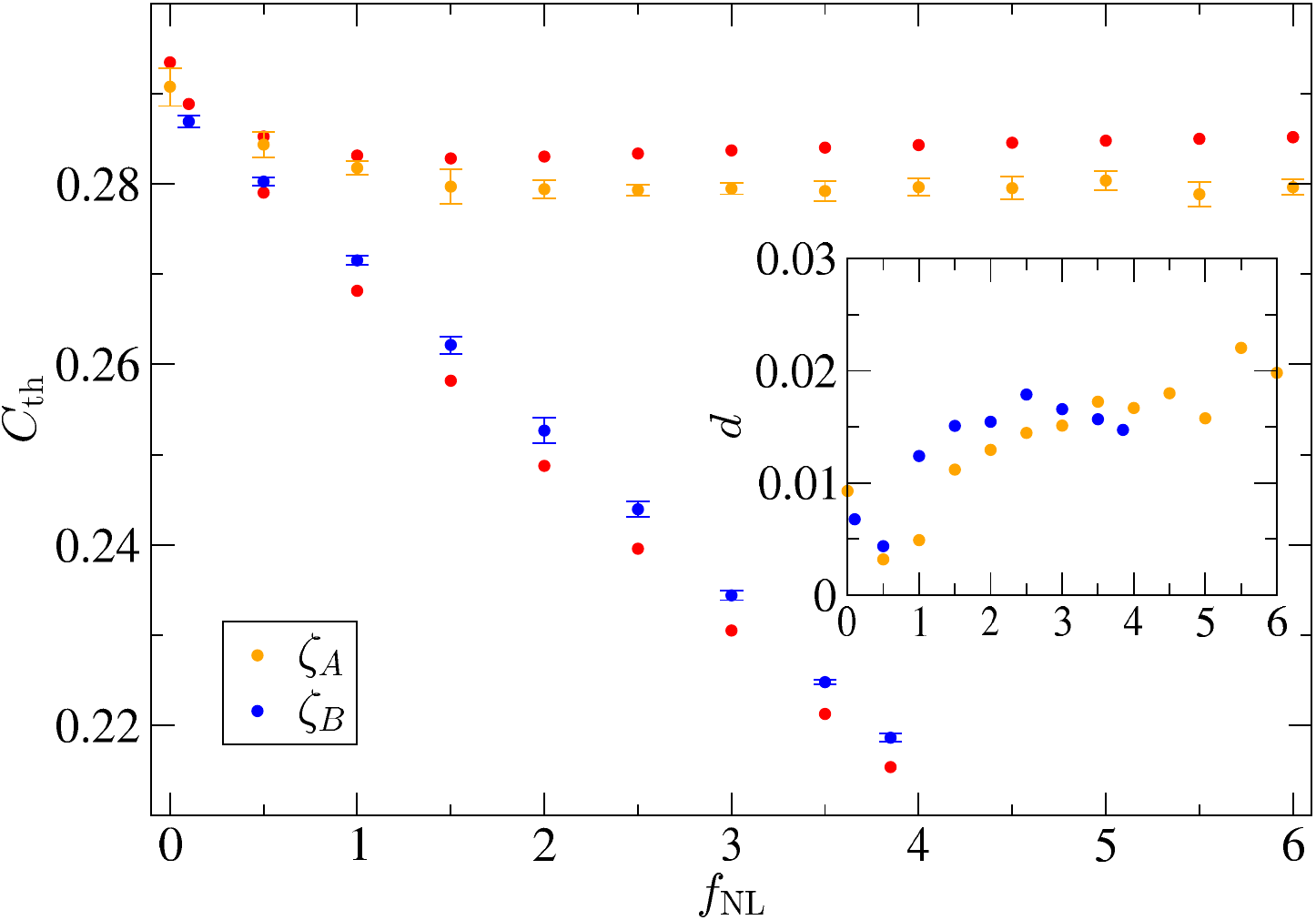}
\caption{Results with  $\Delta \zeta=0$. The orange and blue points represents the values got using the perturbative $\zeta_{A}$ and the non perturbative template $\zeta_{B}$ with the corresponding error bars. The red points are those computed using the universal law of (\ref{eq:universal}). The inner plot represents the deviation $d=\mid \mu_{\rm th}^{N}-\mu_{\rm th}^{A} \mid/\mu_{\rm th}^{N}$ between the numerical $\mu_{\rm th}^{N}$ and the analytical values $\mu_{\rm th}^{A}$ (the same is applied for $\mathcal{C}_{\rm th}$). We also show in dashed line the critical amplitude $\zeta_*\equiv \mu_*$, such that a perturbation jumps into the false local minimum of the potential. For values of $\fnl\sim 3-4$, the thresholds for collapse approaches this limit. \emph{left)} Variation of the threshold for the amplitude $\mu_{th}$  with respect to the non-Gaussian parameter $\fnl$.  \emph{right)} Variation of the threshold for the maximum of the compaction function $\mathcal{C}_{\rm th}$ with respect to the non-Gaussian parameter $\fnl$.}
\label{fig:thresholds1}
\end{figure}

\begin{figure}
\centering
\includegraphics[scale=0.5]{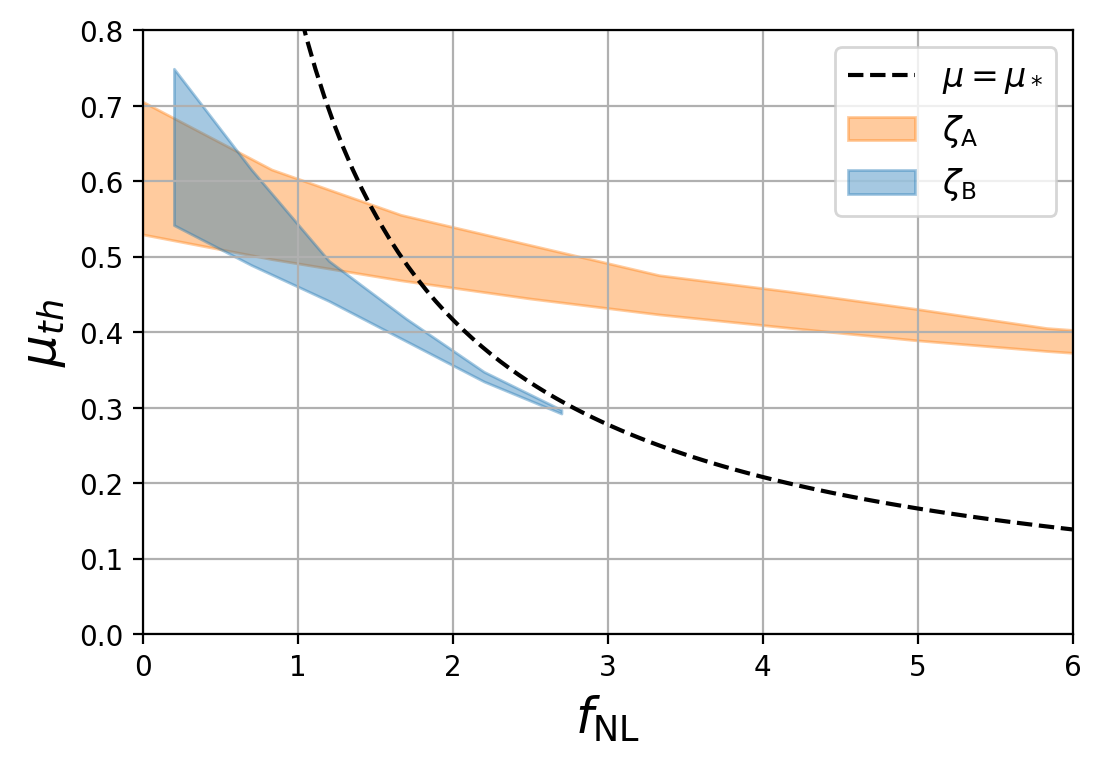}
\includegraphics[scale=0.5]{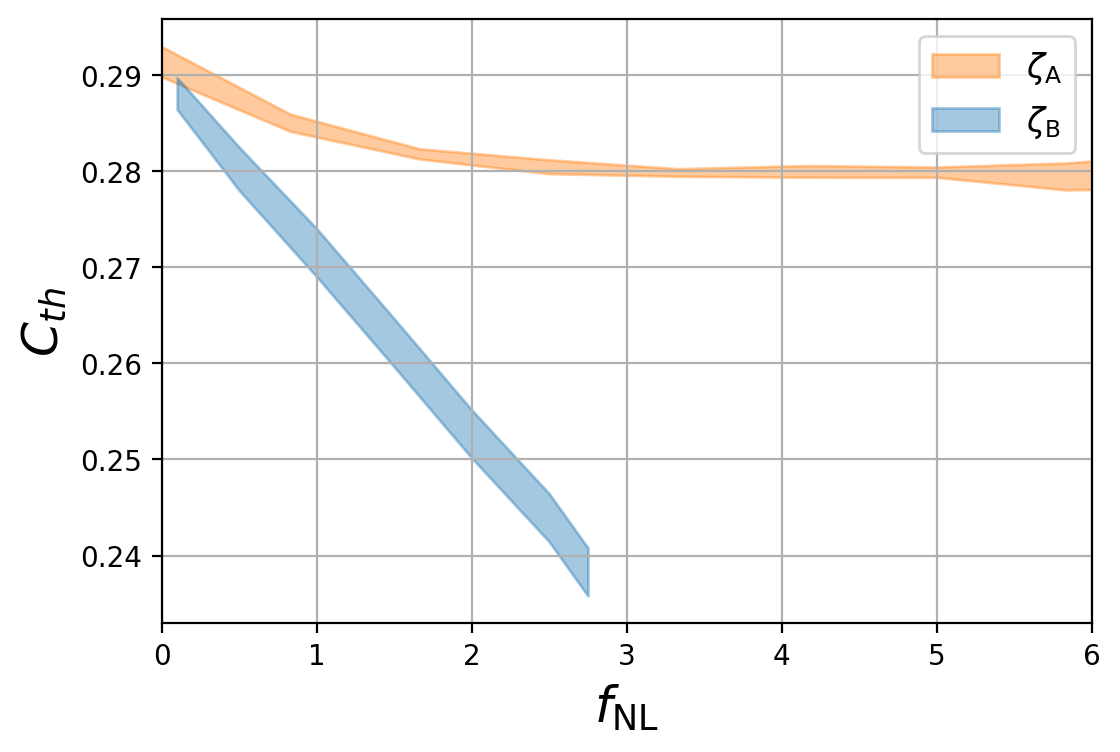}
\caption{Results with $\Delta\zeta \neq 0$ including the dispersion term of (\ref{eq:deltagaus}). Here, we use the numerical value $\nu=\mu/\sigma_0 =5$. \emph{left)} Variation of the threshold for the amplitude $\mu_{th}$  with respect to the non-Gaussian parameter $\fnl$, for both the perturbative template $\zeta_{A}$ (orange) and the non perturbative template $\zeta_{B}$ (blue). The shaded region indicates the dispersion in the numerical results from the dispersion of shapes. \emph{right)} Variation of the threshold for the maximum of the compaction function $\C_{th}$ with respect to the non-Gaussian parameter $\fnl$. While for the perturbative template, the threshold for the compaction function is constant for large $\fnl$, for the non perturbative template the threshold keeps evolving with increasing $\fnl$. In both cases the dispersion in $\C_{th}$ is very small and comparable to the numerical errors.}
\label{fig:thresholds2}
\end{figure}

We have determined the threshold amplitude $\mu_{th}$ and the threshold compaction function $\mathcal{C}_{th}$ for different values of the non-Gaussianity parameter 
in the range $0\leq \fnl\leq 6$.
The numerical results are shown in Fig.~\ref{fig:thresholds1} and Fig.~\ref{fig:thresholds2}. 

In particular, we find that the threshold for the compaction function reaches a constant as we increase the non-linear parameter $\fnl$. 
To gain some insight into the origin of this behaviour, let us note that at sufficiently large $\fnl$ the overdensity is dominated by the non-linear term. 
Indeed, for
\begin{equation}
\mu \fnl \gg 1/\psi(r_m)\approx1.85, \label{approxlargef}
\end{equation}
the median shape can be approximated as
$\zeta(r) \approx \fnl \mu^2 \psi^2(r)$ out to the radius $r_m$.
In the last step in (\ref{approxlargef}) we use $\psi(r) = \sinc(r)$, and $r_m\approx 1.8$ is the maximum of the compaction function for the profile $\psi^2(r)$. In this regime, the shape of the perturbation is independent of $\fnl$, and hence, we expect $\C_{th}$ to be independent of $\fnl$: 
\begin{equation}
\C_{th} \approx 0.286. \quad (\mu_{th} \fnl \gg 2) \label{constantcth}
\end{equation}
Here the numerical value is calculated by evolving the profile $\zeta \propto \sinc^2(k_0 r)$.
From the right panel in Fig. \ref{fig:thresholds1} we see that $\C_{th}$ is indeed nearly constant for $\fnl$ in the range from 2 to 6. This is, however, somewhat coincidental, since  the condition $\mu_{th} \fnl \gg 2$ is only satisfied for $\fnl\gg 10$. 
In the same regime, from $r_m \zeta'(r_m;\mu_{th}))=\sqrt{1-3\C_{th}}-1$, we expect
\begin{equation}
\mu_{th} = \left[{\sqrt{1-3\C_{th}}-1 \over 2 r_m \psi(r_m)\psi'(r_m)}\right]^{1/2} \fnl^{-1/2} \approx 1.12 \fnl^{-1/2}.\quad (\fnl \gg 10)
\end{equation}
Note that this overestimates the actual values of $\mu_{th}$ in the interval $1<\fnl<6$, by 30\% or so (See Fig. \ref{fig:thresholds1}).
The reason is that for $\fnl \lesssim 10$, the value of $r_m$ and, more importantly $\psi(r_m)\psi'(r_m)$, changes appreciably with $\fnl$.  

By contrast with ${\mathcal C}_{th}$, we find that the threshold amplitude $\mu_{th}$ decreases quite significantly with $\fnl$ in the $0<\fnl<6$ interval. 
We also note that the dispersion of the shapes accounts for a very small dispersion of $\C_{th}$.  On the other hand, the threshold for the amplitude $\mu_{th}$  has a larger variability, in particular at low $\fnl$. This may have a sizable impact on the abundance of PBH, although a precise determination of this effect would require simulations which include departures from spherical symmetry (see footnote \ref{sphericalfoot}). Note that for a monochromatic spectrum, the only spherically symmetric profile with finite amplitude at the origin is is precisely the median profile $\zeta_g=\bar\zeta = \mu \sinc(k_0 r)$, so there is no dispersion in the thresholds unless the assumption of spherical symmetry is dropped. In this sense, our treatment of the dispersion by considering the profiles $\zeta^{\pm}$ is only indicative, since it ignores the effect of non-sphericity, which is expected to shift the threshold to slightly higher values.

Recently, the effect of non-Gaussianity with the quadratic template  (\ref{eq:zeta_local_trans}) was also considered in Ref. \cite{Yoo:2019pma,Kehagias:2019eil}, by considering somewhat different approaches. In \cite{Yoo:2019pma}, a fiducial value $\C_{th} \approx 0.267$ was used independently of the value of $\fnl$, and it was concluded that the abundance of PBH grows with $\fnl$. Here, we find that $\C_{th}\gtrsim 0.286$ for any $\fnl$. Note also that the dependence of $\C_{th}$ on $\fnl$ in the range $0<\fnl\lesssim 2$ tends to further enhance the abundance of PBH with growing $\fnl$, relative to the Gaussian case. 

For $\fnl=0$, our result for $\C_{th}$ corresponding to the median profile $\bar\zeta$ coincides with the result reported in \cite{Kehagias:2019eil}, indicating the mutual consistency of the numerical methods. It should be noted, however, that there are some differences in the two approaches, and in the questions we are addressing.  Ref. \cite{Kehagias:2019eil} develops a perturbative method in order to calculate the {\em average} profile for the density contrast $\delta\rho$, where $\delta\rho$ is expanded in powers of $\zeta$ and the calculation is carried out to second order in $\zeta$. Here we consider instead a family of profiles for the curvature perturbation $\zeta$, within a standard deviation from the {\em median} at fixed $\nu$. Ref. \cite{Kehagias:2019eil} finds a value of $\C_{th}$ for the average profile $\langle\delta\rho\rangle$ which is significantly smaller than what we find for the median. This difference is of order $10\%$, for all values of $\fnl$, and it is natural to ask whether this may be due to the difference between the average and the median.
Although these two can indeed be different, we expect the former to be within a standard deviation from the latter, corresponding to 68 \% of all realisations. However, as shown in the right panel of Fig. \ref{fig:thresholds2}, the dispersion of $\C_{th}$ between the profiles $\zeta^{\pm}$ is very narrow, of the order of $1\%$, which is much smaller than the $10\%$ difference mentioned above. A more plausible origin for the discrepancy may be a certain inaccuracy of the perturbative approach used in  \cite{Kehagias:2019eil}, for which the expansion parameter is the amplitude of the curvature perturbation, $\mu\sim 1$. Since this is not small, the accuracy of the truncated expansion is not under control.\footnote{The lack of a small expansion parameter was already noted by the authors of  \cite{Kehagias:2019eil}. They also pointed out that the shape of the second order correction to the average $\delta\rho$ is very similar to that of the lowest order linear term, and that if all subsequent terms were to have a similar profile, then the truncated result would be similar to the fully resummed average profile. Although this remains a logical possibility, which could be checked by calculating further terms in the expansion, it would be surprising to us if this turns out to be the case. As noted above, this would mean that the average profile is several standard deviations away from the median, and therefore far from typical in the ensemble of all realizations. Assuming, for the sake of argument, that this is the case, one should then question what is the point of focussing on the average profile, as opposed to a more representative sample of all realizations.}


\subsection{Case B: Non-perturbative template}


Let us now consider the single field model where the background inflaton overshoots a barrier in the slope of the potential, as in Fig. \ref{potential}. In this case the, the curvature perturbation is related to the Gaussian variable through the non-perturbative relation (\ref{eq:zetanptransf}). Note that the non-perturbative template can also be written in terms of $\fnl$, since $\mu_*$ is a simple function of it, given by eq. (\ref{relation}).

For $\fnl \ll 1$ we expect the results of Case B to be very similar to Case A, and indeed this can be seen in Figs. \ref{fig:thresholds1} and \ref{fig:thresholds2}. Even though the power spectra are slightly different in both cases, the Dirac delta spectrum seems a good approximation to the sharp spike (\ref{eq:model_pw}) which follows from 
the one-field model. 

In Figs. \ref{fig:thresholds1} and \ref{fig:thresholds2} we also plot the curve $\fnl \equiv 5/(6\mu_*)$ as a dashed line. Note that for $\zeta_g \sim \mu_*$ non-linearities are very important, and in fact for $\zeta_g > \mu_*$ the backward fluctuation in the inflaton potential causes a horizon sized region to remain stuck in the false vacumm \cite{Atal:2019cdz}. 
We find that the threshold $\mu_{th}$ for adiabatic perturbations to collapse into PBHs approaches the critical value $\mu_*$ for $\fnl\sim 3.5$. Around this value of $\fnl$, black holes will actually be more likely to be formed though to the creation of false vacuum regions than by adiabatic perturbations. Indeed, we can calculate the abundance of black holes produced by the latter mechanism
\be\label{eq:ab_st_press}
\beta_{st}\propto\int_{\mu_{th}}^{\mu_*} \mu^{3} e^{-\mu^2/ (2\sigma_0^2)} d\mu
\ee
relative  to the abundance of black holes with a baby universe in their interior, given by
\be
\beta_{fv}\propto\int_{\mu_*}^{\infty} \mu^{3} e^{-\mu^2/ (2\sigma_0^2)}  d\mu.
\ee
where we have used the peak theory prescription for computing number density of high peaks \cite{Bardeen:1985tr}, for $\nu=\mu_{th}/\sigma_0\gg1$.
In the same limit, their ratio is then simply given by
\be
\frac{\beta_{st}}{\beta_{fv}}\approx{\mu^2_{th}\over \mu_{*}^2}
\exp\left[\frac{(\mu_{*}^2-\mu^2_{th})\nu^{2}}{2 \mu_{th}^2}\right]-1,
\ee
where we have also used the fact that $\mu_{th}<\mu_*$.
\begin{figure}
\centering
\includegraphics[scale=0.45]{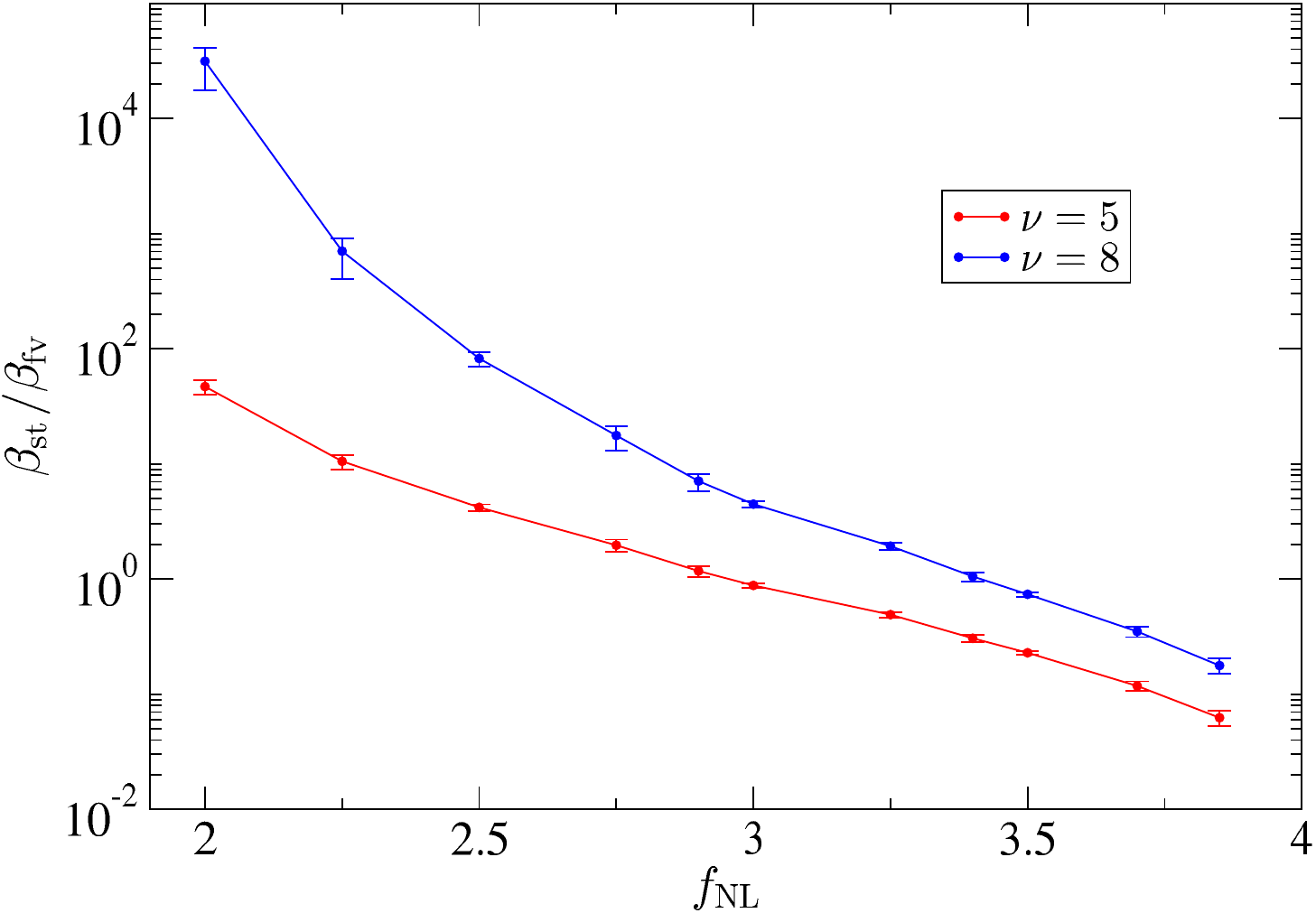}
\caption{Ratio of PBHs coming from the collapse of large overdensities to those created from inflating regions trapped in the false minimum of the potential.}
\label{fig:beta_ratio}
\end{figure}

Note that PBHs created from large overdensities follow the critical collapse scaling, and therefore their mass can range from zero up to the mass contained within the horizon at the time of their formation. On the other hand, PBHs formed from false vacuum bubbles will have a mass which is a fixed (order one) fraction of the mass of radiation contained within a horizon sized region \cite{Garriga:2015fdk}. The ratio (\ref{eq:ab_st_press}) is then an upper bound on the dark matter fraction in the form of standard PBHs relative to that in the form of PBHs containing a baby universe. In Fig. \ref{fig:beta_ratio} we show this ratio as a function of the non-Gaussian parameter $\fnl$, for different values of $\nu$. We see that for $\fnl < 3$ standard black holes dominate, for $3< \fnl < 4$ both types of black holes are produced with a comparable abundance and for $\fnl > 4$, black holes with baby universe in their interior dominate. In principle, as mentioned above, both populations could be distinguished if we could measure the mass distribution of PBH accurately enough to tell whether it follows the critical collapse distribution or it is instead very monochromatic. Whether this can be done realistically is an interesting open question.

\section{Summary and conclusions}

In this paper we have investigated the effect of non-Gaussianities, and of the statistical dispersion in the shape of high peaks, on the threshold for PBH formation.

We assume that the fluctuations $\delta\phi$ of the inflaton field at the time of horizon crossing are Gaussian distributed, so that $\zeta_g$ given in Eq. (\ref{Gaussian}) is a Gaussian random field. This variable is non-linearly related to the standard curvature perturbation through a local relation $\zeta=\zeta(\zeta_g)$. In cosmological perturbation theory, where $\zeta$ is typically very small, it is customary to expand the local relation to second order in $\zeta_g$. The parameter $\fnl$ is then defined as the coefficient of the quadratic term. However, in the context of PBH formation, the curvature perturbation $\zeta$ is sizable, and it is important to consider the full non-perturbative relation between $\zeta$ and $\zeta_g$. In particular, this reveals 
a new regime for PBH formation through the collapse of false vacuum bubbles. These formed at places where a large fluctuation prevented the inflaton from overshooting a small barrier on the slope of the potential \cite{Atal:2019cdz}.

For the evolution of large adiabatic perturbations, we have used the numerical code developed in \cite{albert}. We have investigated the threshold amplitude $\mu_{th}$ for the curvature perturbation to trigger gravitational collapse, and the corresponding threshold $\C_{th}$ in terms of the compaction function, in two different scenarios. In Case A, we used the standard template for perturbative non-Gaussianity, parametrized by $\fnl$, and a monochromatic power spectrum. Case B is based on a more realistic scenario where the inflaton overshoots a barrier, and we use the non-perturbative template for non-Gaussianity. The results of numerical evolution have also been compared successfully with the universal threshold (\ref{eq:universal}), for a broad range of $\fnl$. Both methods agree within a deviation of $\sim 2\%$ as was reported in \cite{EGS}. For the median profiles, the results of this comparison are plotted in Fig. \ref{fig:thresholds1}. 

The results which include the dispersion of shapes are summarized in Fig. \ref{fig:thresholds2}. We find that the effect of the dispersion of shapes on $\C_{th}$ is small, while it is larger on the threshold for the amplitude of fluctuations $\mu_{th}$, particularly at low $\fnl$.  We find that the impact of non-Gaussianity on the thresholds is more substantial in the non-perturbative treatment. For instance, while $\C_{th}$ saturates to a constant 
for  $\fnl \gtrsim 1$ in the perturbative template, we find that in the non-perturbative template it decays approximately linearly as $\C_{th} \approx 0.29 - (0.03 \fnl)$ for $\fnl \lesssim 3.75$.
Numerically, it is hard to probe larger values of $\fnl$ because $\mu_{th}$ approaches $\mu_*$, and the
profiles become extremely peaked near the origin. Nonetheless, we expect
the linear behaviour to saturate to its lowest possible value $\C_{th}\approx 1/5$ for 
$\fnl \gtrsim 4$.

The total effect of non-Gaussianity in the abundances can be inferred directly from the left panel of Fig. \ref{fig:thresholds1}. For the perturbative template  $\mu_{th}$ changes by roughly a factor of $0.5$ as $\fnl$ varies from $0$ to $6$, while for the non perturbative template it changes by a factor of $0.2$ (by extrapoling the curve to $\fnl=6$). Since the abundance of PBHs is exponential in $\mu^2/\sigma_0^2$, it follows that for the larger values of $\fnl$ that we have considered the power spectrum can be a factor of roughly $4$ or $15$ times smaller than it is for the Gaussian case for the perturbative and non-perturbative templates respectively.

The dispersion of $\mu_{th}$ in the ensemble of all realizations of the random field will also have an effect on the determination of the abundances. Here we have estimated such dispersion, illustrated by the shaded regions in Fig.~\ref{fig:thresholds2}, by using the spherically symmetric profiles $\zeta^{\pm}$, which are the envelope of all realizations at one standard deviation from the median. A more precise determination of the dispersion requires the development of numerical codes which can handle non-spherically symmetric realizations in the ensemble. We leave a more detailed consideration of such effect for future work.

Finally, we have computed the relative abundance of PBHs coming from the normal collapse of an overdensity with respect to those coming from false vacuum regions. We conclude that false vacuum regions dominate the production of PBHs for $\fnl\gtrsim 3.5$. PBHs created from large overdensities have a distribution of masses which follows from the critical collapse scaling and the dispersion in shapes, whereas those created from false vacuum bubbles have a fairly monochromatic spectrum. Prospects for observational discrimination of these two possibilities remain an interesting direction for further reseach. Another possible phenomenological application of our results may be in the study of gravitational waves induced by non-Gaussian scalar perturbations \cite{Garcia-Bellido:2017aan,Cai:2018dig,Unal:2018yaa}.

\begin{acknowledgments}
We thank Cristiano Germani and Ilia Musco for valuable discussions. We also thank Nicola Bellomo, Guillem Dom\`enech and Chulmoon Yoo for insightul comments on the manuscript. VA and JG are supported by FPA2016-76005 -C2-2-P, MDM-2014-0369 of ICCUB (Unidad de Excelencia Maria de Maeztu), AGAUR2017-SGR-754. AE is supported by FPA2016-76005-C2-2-P and by the Spanish MECD fellowship FPU15/03583.
\end{acknowledgments}

\end{document}